\documentclass[sigconf]{acmart}
\usepackage[utf8]{inputenc}

\title{Prompt-tuned Code Language Model as a Neural Knowledge Base for Type Inference in Statically-Typed Partial Code}

\author{Qing Huang}
\email{qh@whu.edu.cn}
\affiliation{%
\institution{Jiangxi Normal University}
\country{China}
}

\author{Zhiqiang Yuan}
\email{yuanzhiq@jxnu.edu.cn}
\affiliation{%
\institution{Jiangxi Normal University}
\country{China}
}

\author{Zhenchang Xing}
\email{zhenchang.xing@data61.csiro.au}
\affiliation{%
\institution{CSIRO's Data61 \& Australian National University}
\country{Australia}
}

\author{Xiwei Xu}
\email{Xiwei.Xu@data61.csiro.au}
\affiliation{%
\institution{CSIRO's Data61}
\country{Australia}
} 

\author{Liming Zhu}
\email{Liming.Zhu@data61.csiro.au}
\affiliation{%
\institution{CSIRO's Data61 \& School of CSE, UNSW}
\country{Australia}
} 

\author{Qinghua~Lu}
\email{Qinghua.lu@data61.csiro.au}
\affiliation{%
\institution{CSIRO's Data61}
\country{Australia}
}

\usepackage{balance}
\usepackage{makecell}
\usepackage{graphicx}
\usepackage{float}
\usepackage{array}
\usepackage{tabularx}
\usepackage{diagbox}
\usepackage{multirow}
\usepackage{amssymb}
\usepackage{threeparttable}

\usepackage{setspace}
\usepackage{color}
\usepackage{colortbl}
\usepackage{xcolor}
\usepackage{soul}
\usepackage{booktabs}
\usepackage{verbatim}
\usepackage{enumitem}

\definecolor{mygray}{gray}{.9}
\definecolor{mypink}{rgb}{.99,.91,.95}
\definecolor{mycyan}{cmyk}{.3,0,0,0}
\definecolor{myyellow}{RGB}{255,230,204}
\definecolor{mybule}{RGB}{218,232,252}
\definecolor{mygreen}{RGB}{213,232,212}
\definecolor{titleColor}{RGB}{102,102,102}

\copyrightyear{2022} 
\acmYear{2022} 
\setcopyright{acmcopyright}\acmConference[ASE '22]{37th IEEE/ACM International Conference on Automated Software Engineering}{October 10--14, 2022}{Rochester, MI, USA}
\acmBooktitle{37th IEEE/ACM International Conference on Automated Software Engineering (ASE '22), October 10--14, 2022, Rochester, MI, USA}
\acmPrice{15.00}
\acmDOI{10.1145/3551349.3556912}
\acmISBN{978-1-4503-9475-8/22/10}


\begin{abstract}
Partial code usually involves non-fully-qualified type names (non-FQNs) and undeclared receiving objects.
Resolving the FQNs of these non-FQN types and undeclared receiving objects (referred to as type inference) is the prerequisite to effective search and reuse of partial code.
Existing dictionary-lookup based methods build a symbolic knowledge base of API names and code contexts, which involve significant compilation overhead and are sensitive to unseen API names and code context variations.
In this paper, we formulate type inference as a cloze-style fill-in-blank language task.
Built on source code naturalness, our approach fine-tunes a code masked language model (MLM) as a neural knowledge base of code elements with a novel ``pre-train, prompt and predict'' paradigm from raw source code.
Our approach is lightweight and has minimum requirements on code compilation.
Unlike existing symbolic name and context matching for type inference, our prompt-tuned code MLM packs FQN syntax and usage in its parameters and supports fuzzy neural type inference.
We systematically evaluate our approach on a large amount of source code from GitHub and Stack Overflow.
Our results confirm the effectiveness of our approach design and the practicality for partial code type inference.
As the first of its kind, our neural type inference method opens the door to many innovative ways of using partial code. 
\end{abstract}

\begin{document}

\maketitle

\section{INTRODUCTION}
\begin{figure}
    \centering 
     \includegraphics[width=0.4\textwidth]{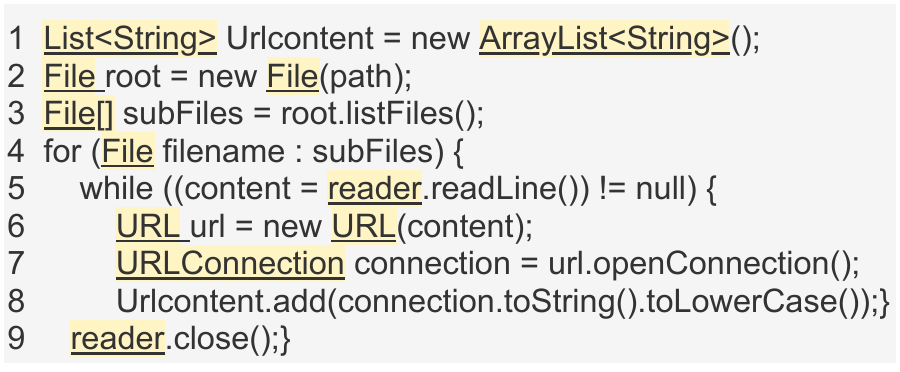}
    \caption{A Partial Code Snippet in Java }
    \label{fig:codeSnippet}
\end{figure}

\begin{figure}
    \centering 
     \includegraphics[width=0.4\textwidth]{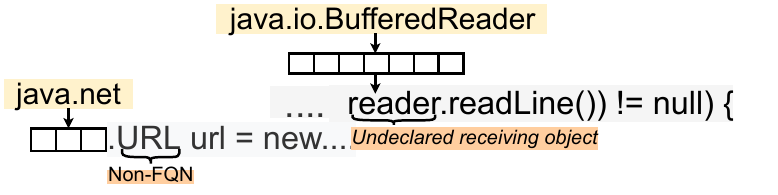}
    \caption{Type Inference as a Fill-in-blank Task} 
     
    \label{fig:URL example}
\end{figure}

Partial code is prevalent in API documentation and online blogs.
Figure~\ref{fig:codeSnippet} shows a typical example of partial code in statically-typed programming language like Java.
Partial code is usually syntactically incomplete.
For example, it may miss the enclosing class and method declaration.
Adding the class and method declaration can easily fix the issue~\cite{Saifullah2019LearningFE}.
The more challenging issue in partial code is that it may invoke methods or access fields on some undeclared receiving objects (e.g., \textit{reader}) or use types by non-fully-qualified names (non-FQNs) (e.g., \textit{List<String>}, \textit{new File()}, \textit{URL}).
Undeclared receiving objects and non-fully-qualified type names will result in the compilation error ``symbol cannot be resolved''.

Effectively parsing partial code and leveraging its knowledge is important for software engineering~\cite{Gupta2020JCoffeeUC,thummalapenta2007parseweb}.
Without resolving undeclared receiving objects and non-FQN types, partial code can only be used as text.
This greatly limits the application of program analysis techniques to partial code, and affects the effective use of partial code.
For example, \textit{URL} in Figure~\ref{fig:codeSnippet} should be \textit{java.net.URL}, not \textit{com.google.gwt.http.client.URL}.
Without knowing the precise type, a code search engine may recommend irrelevant code examples~\cite{Maji2021DCoMAD, thummalapenta2007parseweb}.
Furthermore, several studies show online code snippets often contain API misuses or even malicious behaviors~\cite{Zhang2018AreCE, Piccolboni2021CRYLOGGERDC}.
However, existing code vulnerability analysis tools~\cite{Zhou2019DevignEV, Ren2020APIMisuseDD,TypeInferenceforC} cannot effectively work without accurate type information.

To infer fully-qualified type names (FQNs) for undeclared receiving objects and non-FQN types in partial code, all existing techniques~\cite{Baker, Saifullah2019LearningFE, SnRConstraint} adopt a dictionary-lookup strategy, which uses a symbolic knowledge base that maps simple names and code contexts to FQNs.
The mappings between simple names and FQNs of API elements can be built by parsing library source code or documentation.
However, building the mappings between code contexts and FQNs requires to successfully compile code contexts in which API elements are used.
This compilation overhead limits the amount of collected code contexts and used APIs.
This limitation, together with symbolic name and context matching, results in out-of-vocabulary (OOV) failures when looking up the dictionary with unseen API names or different code contexts.
Although we can theoretically overcome the OOV failures by building more comprehensive knowledge base from more software projects, the compilation overhead makes this solution unrealistic.

As illustrated in Figure~\ref{fig:URL example}, we innovatively formulate type inference as a \textit{cloze-style fill-in-blank language task}.
Instead of relying on compilable software projects to construct symbolic knowledge base for type inference, we fine-tune a masked language model (MLM) pre-trained on source code~\cite{Feng2020CodeBERTAP, kanade2020learning} as the neural knowledge base.
Based on source code naturalness~\cite{Devanbu2012OnTN,Allamanis2018ASO}, language models (including recent large pre-trained language models~\cite{devlin2018bert,Brown2020LanguageMA, Raffel2020ExploringTL}) have been successfully adopted on source code in software engineering tasks~\cite{Feng2020CodeBERTAP, Liebman2010CuebertAN, Allamanis2015BimodalMO,Nguyen2013LexicalSM,Haiduc2010OnTU,Hellendoorn2020GlobalRM}. 
Some recent NLP works show that pre-trained language models can serve as a neural knowledge base of real-world entities and relations~\cite{petroni2019language}, as opposed to symbolic knowledge bases~\cite{Bollacker2008FreebaseAC,Redmon2016YouOL}.
Our work shows for the first time that pre-trained code model has this knowledge base capability for code elements and relations.
Unlike discrete representation of API names and code contexts in symbolic knowledge base~\cite{Baker, Saifullah2019LearningFE, SnRConstraint}, the MLM-based neural knowledge base packs complex code information in a continuous space in the parameters of neural network.

However, the vanilla pre-trained code MLM cannot support the type inference task, because it is trained purely on source code without deep understanding of code syntax and semantics~\cite{AnalyzeCodeBert}.
We design a novel prompt learning method to \textit{stimulate pre-trained code MLM to recognize the form of FQNs and the relationships between API names and code contexts in source code.}
Prompt learning supports ``self-upgrade'' of pre-trained MLM, as pre-training and prompt learning are homogeneous at the MLM core.
We automatically annotate a small amount of compilable library source code with FQNs.
Different from manually defined, uniform prompts in NLP tasks~\cite{Sun2021NSPBERTAP,Han2021PTRPT,Gu2021PPTPP,Ding2021PromptLearningFF,Liu2021PTuningVP}, our approach automatically extract code lines surrounding a focused code line with FQNs as its contextual FQN prompt.
We mask the FQNs in the focused code line and train the MLM to predict the masked FQNs in the FQN prompts.
Instead of commonly used random masking~\cite{devlin2018bert,Feng2020CodeBERTAP}, we design full span masking strategy that fits with the FQN characteristics and our type inference task.
Although FQN prompts contain only local code contexts, the model learns long-range dependencies among code elements through FQN prompts with overlapping code lines.

Our type inference task aligns perfectly with the pre-training and prompt learning of code MLM.
Given an unseen partial code, we form code prompts with masks for non-FQN types and undeclared receiving objects in the same form as the FQN prompts used for prompt learning, and use the prompt-tuned code MLM to fill in the masks with the missing FQNs.
In NLP tasks, prompts use a fixed-length mask span.
However, FQNs vary in length.
We develop a variable-length mask prediction method to infer the variable-length FQNs by searching for the FQN length with maximum probability.
Unlike symbolic name and context matching in existing methods~\cite{Saifullah2019LearningFE,SnRConstraint,Phan2018StatisticalLO}, our inference uses code prompt to activate the FQN knowledge packed in the parameters of the code MLM.

Our evaluation focuses on two aspects, i.e., effectiveness and practicality. 
For effectiveness, we evaluate how intrinsic factors (prompt learning data size and FQN masking strategy) affect our model performance and find that prompt learning with full span masking strategy on 10\% of library source code can significantly boost the type inference accuracy, surpassing the vanilla pre-trained code MLM by an increase of 0.539. 
Then we evaluate how extrinsic factors (code context similarity and API cardinality) affect our model and find that our model makes accurate type inference in face of large context variations and high API cardinalities.

For practicality, we evaluate the model performance on partial code snippets from Stack Overflow. 
Our model outperforms the state-of-the-art tools COSTER~\cite{Saifullah2019LearningFE} and SnR~\cite{SnRConstraint} (on average accuracy 0.91 versus 0.71 and 0.87).
Furthermore, for the FQNs and code contexts unseen during prompt learning, our model still achieves the accuracy around 0.7. 

Our approach is the first successful attempt to adopt the ``pre-train, prompt and predict'' paradigm~\cite{Raffel2020ExploringTL,Brown2020LanguageMA,Schick2021ItsNJ, Schick2021ExploitingCF} to support program analysis in software engineering. 
The outstanding performance of our model is rooted in the perfect alignment of large code pre-training, effective FQN prompt learning, and type inference as a fill-in-blank task.
This alignment not only allows FQNs and code contexts to be effectively embedded in a neural code knowledge base, but also allows FQNs (even unseen ones) to be accurately inferred from variant code contexts.

In this paper, we make the following contributions:
\begin{itemize}[leftmargin=*]
    \item We are the first to formulate the type inference problem as a cloze-style fill-in-blank language task, and propose a novel ``pre-train, prompt and predict'' paradigm to solve the task.

    \item We design the first neural partial code type inference approach that does not rely on symbolic knowledge base of API names and code contexts. In contrast, our approach uses a prompt-tuned code MLM as a neural knowledge base which makes it realistic to overcome the OOV failures in symbolic type inference. 
    
    \item We design the first prompt learning method to stimulate task-agnostic pre-trained code MLM to recognize fine-grained FQN syntax and usage, and are the first to apply prompt-tuned code MLM for the partial code type inference task.
    
    \item Driven by code-specific task needs, we design automatic and contextual FQN prompt and full-span mask strategy for prompt tuning, and develop variable-length mask prediction method.

    \item We systematically evaluate our approach design. Our approach achieves superior performance even with only a small amount of prompt learning code, and is the first approach with the promising capability of few-shot type inference. Our data package can be found
here\footnote{\href{https://anonymous.4open.science/r/Experiment-Datas-93AE/README.md}{https://anonymous.4open.science/r/Experiment-Datas-93AE/README.md}}. The code will be released upon paper acceptance.
    
    
\end{itemize}

\section{Problem Definition}\label{sec:problem-Def} 

Different from existing dictionary-lookup based type inference methods~\cite{Saifullah2019LearningFE,SnRConstraint, Phan2018StatisticalLO}, we formulate 
the type inference in statically-typed partial code as a \textit{cloze-style fill-in-blank language task}.
Given an input partial code snippet, we make type inference at three types of places (those highlighted in yellow in Figure~\ref{fig:codeSnippet}:
1) data type of variable declaration;
2) type name of class instantiation and array creation; and
3) the object or the type on which a method is invoked or a field is accessed.
Type can be simple type including annotation and enum) (e.g., \textit{URL}), array type (e.g., File[]), and generic type (e.g., List<String>).
We do not consider other variable usage (e.g., variable assignment, method argument, array access), for example \textit{content} in the lines 5 and 6 in Figure~\ref{fig:codeSnippet}, because the relevant data type can be derived from assignment expression and method signature after resolving the missing types in partial code~\cite{dagenais2008enabling}.
We do not make inference for chained method calls and field accesses, for example, \textit{connection.toString().toLowerCase()} in the line 8 in Figure~\ref{fig:codeSnippet} 
Once \textit{connection} is resolved, the receiving type of \textit{toLowerCase()} can be inferred from the return type of \textit{connection.toString()}.
We refer to non-FQN type names (e.g., \textit{File}, \textit{URL}) and undeclared receiving objects (e.g., \textit{reader}) as \textit{type inference points}.
As illustrated in Figure~\ref{fig:URL example}, our goal is to fill in the missing type information at type inference points in the same way as a person completes a fill-in-blank language task.
Although type inference provides the basis for making a partial code compilable, this work does not aim to produce a syntactically correct code.

\section{APPROACH}

\begin{figure}
    \centering 
     \includegraphics[width=0.47\textwidth]{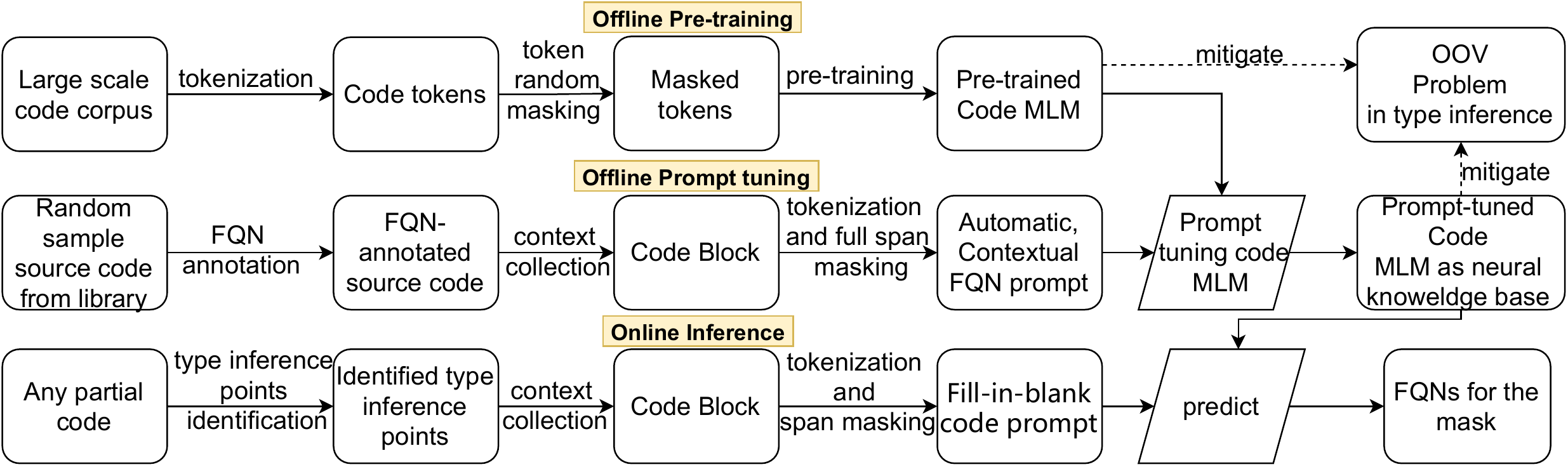}
    \vspace{-4mm}
    \caption{Approach Overview}
    \label{fig:overview}
    \vspace{-5.4mm}
\end{figure}

We propose a novel masked language model (MLM) based ``pre-train, prompt and predict'' paradigm to solve the fill-in-blank task for type inference.
As shown in Figure~\ref{fig:overview}, in offline learning phase, we choose a MLM pre-trained on a large corpus of source code, without the need of code compilation.
Then, we design code-specific prompt learning method which uses a small amount of compiled library source code to stimulate pre-trained code MLM to recognize FQN syntactic and usage patterns.
The offline learning produces a prompt-tuned code MLM whose learning objective aligns perfectly with the fill-in-blank task for type inference.
At the inference time, the type inference point in the partial code is converted into a code prompt with masks, fed into the prompt-tuned code MLM, serving as a neural knowledge base to predict the tokens at the masks.

\subsection{Pre-training MLM of Source Code}\label{sec:PLM}

Exisitng type inference methods~\cite{Saifullah2019LearningFE,SnRConstraint,Phan2018StatisticalLO} suffer from the OOV failures caused by the reliance on compilable code contexts to build symbolic knowledge base of API usage.
To remove this reliance, our approach works on raw source code.
Studies on source code naturalness~\cite{Devanbu2012OnTN, Allamanis2018ASO} show that code can be understood and manipulated in the same way as natural language text.
In this work, we adopt a Transformer-based MLM.
Transformer~\cite{Vaswani2017AttentionIA} is the backbone of the state-of-the-art language models.
A MLM, such as BERT~\cite{Devlin2019BERTPO} is trained to predict the masked word in the input text.
This training objective aligns perfectly with our formulation of type inference as a fill-in-blank language task.

In our current implementation, we use the pre-trained CodeBert model~\cite{Feng2020CodeBERTAP} which was trained on a large code corpus provide by Husain et al.~\cite{Husain2019CodeSearchNetCE}.
We choose CodeBert because many studies~\cite{Karmakar2021WhatDP,Troshin2022ProbingPM,Wan2022WhatDT} confirm that the model captures rich code information and demonstrate its effectiveness in many downstream tasks~\cite{Lu2021CodeXGLUEAM, wang2020detecting,Zhou2019DevignEV,tufano2019empirical}.
However, our approach is not limited to CodeBert, but can use any MLMs.
The training code corpus can be easily expanded with some code crawling effort, as model training does not require any compiled code.
A technical challenge to apply language models to a large code corpus is the rare words of program element names. 
Using a word tokenizer will result in severe out-of-vocabulary tokens and poor representation for rare tokens.
To overcome this issue, we use subword tokenizer WordPiece~\cite{Wu2016GooglesNM} which is a standard practice to deal with the token OOV issues.

\vspace{-2.3mm}
\begin{figure}[h]
    \centering 
     \includegraphics[width=0.47\textwidth]{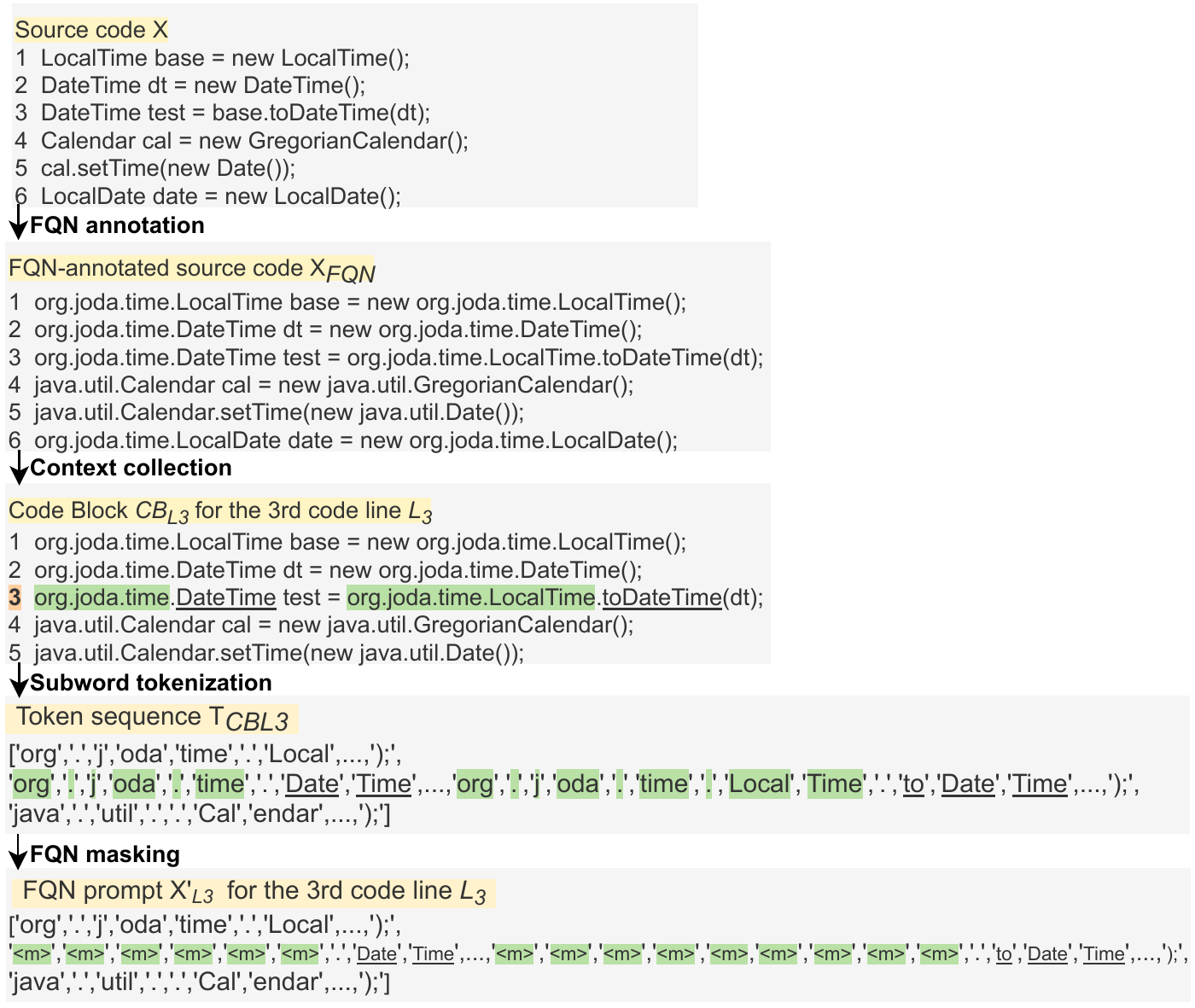}
    \caption{An Example of FQN Prompt Generation }
    \label{fig:fine-tuning-example}
    \vspace{-5.5mm}
\end{figure}

\subsection{Prompt-based FQN Learning}\label{sec:Knowledge activation}

The vanilla pre-trained code MLM does not have deep understanding of code syntax and semantics~\cite{TypeInferenceforC, AnalyzeCodeBert}, because it is trained on raw source code as text.
We propose a prompt tuning method which uses a small amount of FQN stimuli (i.e., prompts) to upgrade the pre-trained MLM with FQN syntax and usage.
Contextual FQN prompts are automatically derived from compiled library source code.
Different from existing work~\cite{Gu2022AssembleFM,Wang2021BridgingPM} that uses heterogeneous downstream tasks to fine-tune the pre-trained MLM, our model pre-training and prompt learning are homogeneous.
Both are MLM learning and align with our type inference task formulation.

\subsubsection{Automatic, Contextual FQN Prompt Generation}\label{sec:promp-design}
We generate FQN prompts from some compiled source code files of the libraries whose APIs need to be inferred.
Compiling a library is much easier than compiling a large number of software projects using the library. We only need a small amount of randomly sampled library source code (e.g., 10\%) to achieve highly effective prompt tuning.
This is because prompt tuning stimulates pre-trained MLM to recognize FQN syntactic and usage patterns in the large code corpus on which it is trained, rather than simply memorizing some FQNs and limited usage contexts used for model fine tuning.

Given the code file $X=\{L_i\}$ ($L_i$ ($1 \leq i \leq n)$ are $n$ code lines in the file), FQN prompt generation $G(X)$ includes a sequence of automatic code transformations: FQN annotation, context collection, subword tokenization, and FQN masking, and outputs $k$ ($k \leq n$) distinct FQN prompts $X'_{L_i}$ for the code lines $L_i$ with FQNs. 
Here, the code lines $L_i$ are separated by the semicolon (i.e., ``;''), not the newline.
This design breaks down the syntactic boundaries between code elements, which treats the code more similar to the inputs of model pre-training and results in long prompts.
Furthermore, splitting code lines by ``;'' brings adjacent more code elements in the same lines, which makes the resulting code lines contain as much FQN information as possible so that MLM learns richer FQN syntax and usage information from an FQN prompt during prompt learning.
Figure~\ref{fig:fine-tuning-example} illustrates the process.
Except for the FQN information, $G(X)$ does not use any other code analysis (e.g., AST or data flow).
We feed the FQN prompt $X'_{L_i}$ to the pre-trained MLM to predict the masked FQN tokens.
The original code token sequence is the ground truth to tune the MLM by minimizing the cross-entropy loss between the ground-truth toke and the prediction (the same learning objective as the MLM pre-training~\cite{devlin2018bert}).

\textbf{FQN annotation}.
We perform FQN annotations at the three types of type inference points defined in Section~\ref{sec:problem-Def}.
As the code $X$ is compilable, annotating it with FQNs is straightforward.
For example, simple name \textit{LocalTime} in line-1 is replaced by its FQN \textit{org.joda.time.LocalTime}.
For the receiving object of method invocation or field access, we replace the variable name with the type FQN of the variable.
For example, \textit{base.toDateTime()} is annotated as \textit{org.joda.time.LocalTime.toDateTime()}.
This abstracts away variable name variations but keeps only the receiving type of method invocation or field access.
Although the code is from a particular library, it may use APIs of other libraries, such as ``Calendar'' in line-4 from the Java SDK, which are also annotated with FQNs.
We denote FQN-annotated code as $X_{FQN}$.

\textbf{Context collection.}
Next, we scan $X_{FQN}$ from top down to collect a code block $CB_{L_i}$ as the local context of each code line $L_i$.
If $L_i$ does not contain FQNs, we ignore this code line.
The code block consists of up to $t$ adjacent code lines before and after $L_i$, denoted as $L^t_p$ and $L^t_s$ respectively, i.e., $CB_{L_i}$=$L^t_pL_iL^t_s$.
In Figure~\ref{fig:fine-tuning-example}, the code block $CB_{L_3}$ for the 3rd code line contains five lines of code: line-1\&2 ($L^t_p$), line-3 ($L_2$) and line-4\&5 ($L^t_s$).
In this work, we set $t=2$, that is, a code block may have up to 5 lines of code, mainly limited by the window size of the MLM (512 tokens in our current implementation).
The benefit of using a small local context is that it forces the model to learn to make inference from limited context information which is often the case for (short) partial code.

\textbf{Subword tokenization.}
We tokenize the code block by the same WordPiece~\cite{Wu2016GooglesNM} tokenizer used for pre-training MLM, and obtain a token sequence $T_{CB_{Li}}=T_{L^t_p} \oplus T_{L_i} \oplus T_{L^t_p}$ for each code block, where $T_{L^t_p}$, $T_{L_i}$ and $T_{L^t_p}$ are the token sequence for $L^t_p$, $L_i$ and $L^t_s$, respectively, and $\oplus$ is sequence concatenation.

\textbf{FQN masking.}
Finally, given the token sequence $T_{CB_{Li}}$ of the code block $CB_{L_i}$, we mask the FQN tokens in $T_{L_i}$, and leave the tokens in $T_{L^t_p}$ and $T_{L^t_s}$ unchanged.
As a result, we obtain $k$ ($k \leq n$) FQN prompts $X'_{L_i}=T_{L^t_p} \oplus mask(T_{L_i}) \oplus T_{L^t_p}$ for the $k$ code lines $L_i$ with FQNs.
As each prompt has a different focused code line with FQN masks, none of the $k$ prompts are identical.
Our FQN prompts have overlapping code lines, which allows the model to learn long-range dependencies beyond local contexts.

$mask(T_{L_i})$ masks all FQNs in $L_i$, and keep all other tokens unchanged.
The FQN tokens that do not appear in the original code $X$ will be masked.
For example, for the FQN \textit{org.joda.time.DataTime} in the $L_3$ in Figure~\ref{fig:fine-tuning-example}, we mask the package name prefix \textit{org.joda.time} before \textit{DataTime}.
For the FQN of the receiving object of the method call \textit{base.toDateTime()}, we mask the whole class FQN \textit{org.joda.time.-LocalTime} before \textit{toDateTime()}.
Our mask method aligns with our fill-in-blank type inference task, and produces FQN prompts that stimulates the MLM to learn to fill in the missing type information.
We mask not only name tokens but also name separators (e.g., ``.'' for Java).
This lets the model learn to generate FQNs in the proper form.
We consider two masking strategies: \textit{random mask} or \textit{full span mask}.
Random mask is the commonly used strategy in NLP~\cite{devlin2018bert,Feng2020CodeBERTAP}.
In our application, we follow the CodeBert setting, i.e., randomly mask 15\% tokens in the FQN parts to be masked. 
Considering the fact that parts of a FQN are integral as a whole, we design the full span mask which masks all the tokens in the FQN parts to be masked.
Studies~\cite{Guu2020REALMRL} show span masking makes the mask prediction more challenging and forces the model to learn better embeddings.

\subsection{Type Inference as a Fill-in-blank Task}\label{sec:type inference}

Given a partial code $X_{pc}$, we use the prompt-tuned code MLM as a neural knowledge base to infer the FQNs of non-FQN type names and undeclared receiving objects in $X_{pc}$.
As partial code cannot be compiled in general~\cite{7486323}, we develop regular expressions similar to those in existing type inference methods~\cite{SnRConstraint, Phan2018StatisticalLO} to identify type inference points based on code lexical patterns.

We perform type inference for one type inference point at a time.
Let $I$ be a type inference point in line $L_j$.
We collect a code block $CB^I_{L_j}$ in the same way as described in Section~\ref{sec:promp-design}.
We use the WordPiece tokenizer to tokenize the code block.
As illustrated in Figure~\ref{fig:URL example}, for a non-FQN name (e.g., \textit{URL} in Figure~\ref{fig:codeSnippet}), we prepend mask tokens to the non-FQN type name.
For an undeclared receiving object (e.g., \textit{reader} in Figure~\ref{fig:codeSnippet}), we replace the variable name with mask tokens.
The resulting token sequence $X^I_{L_j}$ forms a fill-in-blank code prompt, in the same form as the FQN prompts used for tuning the MLM.
$X^I_{L_j}$ is fed into the prompt-tuned MLM to predict the masked tokens, i.e., fill in the missing type names in $X_{pc}$.
The predicted tokens are concatenated to form the FQNs.
In NLP, prompts contain fixed-length mask span.
However, the length of FQNs vary.
We produce multiple prompts $X^I_{L_j}$ with different lengths of masks, and compare the average probabilities of the mask predictions at different lengths to select the best prediction.
We use the minimum and maximum mask span $min$ and $max$, i.e., the shortest and longest subtoken length of the FQNs observed in the libraries. 
$min$=3 and $max$=69 in our current implementation.
It's worth noting that the known FQNs in the generated code prompts have upper and lower bounds.
The upper bound of known FQNs in a code prompt means only the type inference point to be predicted is simple name while the FQNs of all other simple names are known.
This is the easiest case for FQN inference (referred to as leave-one-out).
In contrast, the lower bound means the FQNs of all simple names are unknown.
This is the most difficult case for FQN inference (referred to all-unknown).
In fact, the code prompt generated by the real partial code lies in between the upper and lower bounds, where, in addition to the type inference point to be predicted, the FQNs of some simple names would be known while others may be unknown.

\section{Evaluation}
We evaluate our approach from the two following aspects:
\begin{itemize}[leftmargin=*]
    \item \textbf{Effectiveness}: we investigate how \textbf{intrinsic factors (RQ1)} (i.e., prompt-tuning data size and FQN masking strategy) and \textbf{extrinsic factors (RQ2)} (i.e., prompt similarity and API cardinalities by simple names) affect our approach.

    \item \textbf{Practicality}, we investigate (\textbf{RQ3}) how \textbf{accurate} is our approach for inferring types in code snippets collected from Stack Overflow, (\textbf{RQ4}) how well our approach performs for the \textbf{FQNs and code contexts seen or unseen} during prompt tuning, and (\textbf{RQ5}) what are \textbf{typical failure patterns and causes}.
\end{itemize}
\textbf{Hardware Configuration.} These experiments are conducted with CUDA on NVIDIA GeForce RTX 3090. 
The operating system is Ubuntu 20.04.3 LTS.

\subsection{Effectiveness Evaluation}\label{effective}

\subsubsection{Motivation}
Our approach is the first type inference method based purely on a neural knowledge base encoded in the code MLM.
The FQN prompt learning is affected by two intrinsic factors:
1) the amount of code used for prompt learning;
2) the FQN mask strategy (random mask versus full span mask).
Furthermore, two extrinsic factors also affect the type inference performance:
1) the code similarity between the code prompt for type inference and the FQN prompts for prompt learning;
2) the number of APIs whose simple names are the same (referred to as cardinality or ambiguity for type inference~\cite{Saifullah2019LearningFE}).
It is important to study and understand how these intrinsic and extrinsic factors affect our approach in particular and neural type inference method in general.

\subsubsection{Dataset}
\label{sec:effectivenessdata}
To obtain reliable results, we need a large amount of code snippets with ground-truth FQNs.
We consider six libraries, including \href{https://github.com/hmkcode/Android}{Android}, \href{https://github.com/gwtproject/gwt}{GWT}, \href{https://github.com/hibernate/hibernate-orm}{Hibernate}, \href{https://github.com/JodaOrg/joda-time}{Joda Time}, \href{https://github.com/x-stream/xstream}{Xstream}, and 
\href{https://github.com/openjdk/jdk}{JDK} used in previous type inference work~\cite{Phan2018StatisticalLO,Saifullah2019LearningFE,SnRConstraint}.
Different from previous work that collects a small amount of code using these library APIs, we use the library source code itself for large-scale effectiveness evaluation.
We download the source code from the library's Github repositories, and compile the code successfully.
In total, we collect 39,255 source code files, including 3,190, 7,096, 10,670, 17,404, 330 and 565 source code file from Android, GWT, Hibernate, JDK, Joda Time and Xstream, respectively.
As the focus of this evaluation is on the capability of FQN prompt learning and inference, we collect all methods, not just public API methods in these code files.

We randomly split these 39,255 files into two portions: 40\% as the prompt learning dataset and 60\% as the type inference dataset.
We further randomly split the prompt-learning data into four equal portions.
We use \textit{x-pl} (x=1, 2, 3, or 4) to denote the number of these four portions used for prompt learning.
Note that each prompt tuning involves data from six libraries.
We use the FQN annotation method described in Section~\ref{sec:promp-design} to annotate these source code files for prompt learning and as ground-truth to evaluate type inference results.
From the prompt learning methods, we obtain 361,048 FQN prompts which involve 133,739 distinct FQNs.
At the inference time, we use as input the code prompt to simulate partial code. 
As the input code does not include import and field declarations, the types used in the method become non-FQN types, and the fields used become undeclared receiving objects. 
Furthermore, we randomly remove 0-2 variable declarations to create more undeclared receiving objects.
From the type inference methods, we obtain 541,020 code prompts which involve 183,298 distinct FQNs to be inferred.
71.34\%, 57.18\%, 46.82\% and 38.78\% of these to-be-inferred FQNs do not appear in the \textit{1-pl}, \textit{2-pl}, \textit{3-pl}, \textit{4-pl} prompt learning methods, respectively.
None of the code blocks used for prompt learning and type inference are identical.
Here, we generate the code prompts by the leave-one-out setting (see section~\ref{sec:type inference}), in order to understand the performance upper bound of our approach.

\subsubsection{Metrics}
We use accuracy and BLEU score~\cite{Lin2004ORANGEAM} to measure the performance of our approach. 
The accuracy is the percentage of correctly inferred FQNs.
Unlike existing methods~\cite{SnRConstraint, Saifullah2019LearningFE}, we do not use recall and F1-score, because existing methods may fail to make the inference due to API or context OOV failures, while our approach can always make an inference.
As such, accuracy, recall and F1-score are the same for our approach.
The BLEU score~\cite{Lin2004ORANGEAM} is commonly used for evaluating text generation tasks.
It is based on n-gram matches between the generated text and the reference text.
Here, we calculate the BLEU score to compare the similarity between the predicted FQN and the ground-truth FQN.
The BLEU score is 0.00 -1.00, and the higher the better.
As the shortest FQNs to be inferred contain only three tokens, we compute BLEU-2 (i.e., 2-gram).
We compute BLEU on the subword tokens so that we can understand the matching at a finer-grained level.

\subsubsection{RQ1: Intrinsic Factors}\label{sec:rq1}

This RQ includes the experiments on prompt-learning data size and FQN mask strategy respectively.

\paragraph{Prompt-Learning Data Size}\label{exper:data-size}
We use four sizes of prompt-learning data (\textit{1-pl}, \textit{2-pl}, \textit{3-pl}, \textit{4-pl}) to fine-tune the MLM and test it on the same type inference dataset.
Here we use full span mask strategy.
We refer to the prompt-tuned MLM as $M_{F}-x$ for the \textit{x-pl} portion.
In addition, we use a zero-shot setting to contrast the effectiveness of prompt learning.
In the zero-shot setting, the vanilla code MLM is used directly to infer the FQNs in the test data.

\begin{table*}[h]
    \centering
    \caption{The Effects of Prompt Learning and Full Span Masking Strategy}
    \vspace{-3mm}
    \begin{tabular}{|c|c|c|c|c|c|c|c|c|c|c|c|c|}
    \hline \multirow{2}{*}{Libraries} & \multicolumn{2}{c}{Zero-shot} & \multicolumn{2}{|c|}{$M_{F}-1$} & \multicolumn{2}{c}{$M_{F}-2$} & \multicolumn{2}{|c|}{$M_{F}-3$} & \multicolumn{2}{c|}{$M_{F}-4$} & \multicolumn{2}{c|}{$M_{R}-4$} \\\cline{2-13}
	 & Acc	& BLEU-2 & Acc & BLEU-2 & Acc & BLEU-2 & Acc & BLEU-2 & Acc & BLEU-2 & Acc & BLEU-2 \\\hline
     Android & 0.355	& 0.523 & 0.822 & 0.892 & 0.829 & 0.897 & 0.849 & 0.902 & 0.859 & 0.911 & 0.415 & 0.527 \\\hline
     Gwt & 0.254 & 0.403 & 0.866 & 0.952 & 0.870 & 0.960 & 0.902 & 0.965 & 0.915 & 0.970 & 0.292 & 0.386 \\\hline
     Hibernate & 0.215 & 0.341 & 0.866 & 0.956 & 0.885 & 0.965 & 0.918 & 0.975 & 0.923 & 0.977 & 0.232 & 0.342 \\\hline
     JDK & 0.341 & 0.505 & 0.787 & 0.906 & 0.808 & 0.909 & 0.843 & 0.923 & 0.858 & 0.926 & 0.387 & 0.502 \\\hline
     Joda time & 0.402 & 0.633 & 0.872 & 0.960 & 0.894 & 0.965 & 0.911 & 0.968 & 0.911 & 0.970 & 0.413 & 0.607 \\\hline
     Xstream & 0.274 & 0.465 & 0.863 & 0.939 & 0.839 & 0.939 & 0.904 & 0.958 & 0.900 & 0.954 & 0.430 & 0.555 \\\hline
     Average score & 0.307 & 0.478 & 0.846 & 0.934 & 0.854 & 0.939 & 0.888 & 0.949 & 0.894 & 0.951 & 0.362 & 0.486 \\\hline
    \end{tabular}
    \label{tab:M-x}
    \vspace{-3mm}
\end{table*}

\textbf{Result and analysis.}
Table~\ref{tab:M-x} shows the results.
The vanilla code MLM (without prompt learning) achieves an average accuracy 0.31 and the BLEU score 0.48 in the zero-shot setting.
This shows that even the vanilla code MLM captures some useful FQN information, the information is not sufficient to accurately predict the FQNs.
We find that many correct predictions are short and frequently used FQNs, such as \textit{java.io.File}.
But the vanilla MLM generally fails for relatively complex FQNs.
For example, it predicts the wrong FQN \textit{.animationation.Animator.ObjectAnimator} for \textit{ObjectAnimator}, not the correct \textit{android.animation.Animator.ObjectAnimator}.

When we use 10\% library source code files for prompt learning (i.e., $M_F-1$), the accuracy rises sharply from 0.31 to 0.85, with an increase of 0.54. 
It shows that prompt learning is effective to stimulate the vanilla code MLM to learn FQN syntactic and usage patterns, even with a small portion of library code.
For example, $M_F$-1 correctly predicts \textit{android.animation.Animator.ObjectAnimator} for \textit{ObjectAnimator}.
In addition, the average BLEU-2 is also much higher (i.e., 0.93 vs. 0.48).
This shows that the predicted FQN is very similar to the ground-truth FQN even they are not exactly the same.
With additional tool support (e.g., search engine), one may find the correct FQNs from the almost correct predictions.
For example, searching \textit{android.raphics.Canvas.clipRect()} (missing ``g'' for graphics) on Google will find at the top-1 rank the correct API name.
As 71.34\% FQNs do not appear in the \textit{1-pl} prompt learning data, our results show $M_F$-1 obtains excellent zero-shot inference capabilty at the FQN level after prompt learning.

With the gradual increase of prompt learning data, the average accuracy and BELU-2 score increases more incrementally, compared with the sharp increase from the zero-shot setting to $M_F-1$.
The model performance becomes stable at $M_F$-3 (average accuracy 88.8 and average BLEU-2 score 0.95), with only marginal increase at $M_F$-4 (average accuracy 0.89 and average BLEU-2 0.95.).
This suggests that prompt learning or not is much more fundamental to upgrade the vanilla code MLM than the amount of prompt learning data.

Compared with the other four libraries, JDK and Android achieve relatively low accuracy and BLEU-2 score.
This is because JDK and Android are software development kits which are essentially collections of many libraries.
These SDK libraries have variants FQNs.
Especially for JDK, its FQNs can vary greatly, such as \textit{javax.xml}, \textit{org.w3c.dom}, \textit{org.omg.CORBA}, \textit{org.xml.sax}, etc.
This increases the difficulty in accurately inferring the diverse forms of FQNs in JDK, especially when prompt learning data is small (e.g., $M_F-1$).

\paragraph{Masking Strategy} \label{mask-strategy}
We use the four sizes of prompt-learning data to fine-tune the MLM but we use random masking strategy.
We refer to the prompt-tuned MLM as $M_{R}-x$ for the \textit{x-pl} data portion, which are tested on the same inference dataset as $M_F-x$.

\textbf{Result and analysis}
Table~\ref{tab:M-x} shows the results of $M_R-4$.
Due to the space limitation, we omit $M_R-1$, $M_R-2$ and $M_R-3$ as $M_R-4$ achieves the best results.
Unlike the sharp increase by $M_F-x$, $M_R-4$ achieves only slightly better accuracy than the zero-shot setting (0.36 versus 0.31), and $M_R-4$ and the zero-shot setting achieve the close BLEU-2 scores (0.49 vs. 0.48).
This suggests that prompt learning requires an appropriate masking strategy to achieve its effectiveness.
As the commonly used random masking strategy does not fit with FQN data characteristics and type inference task, it performs poorly.
In contrast, full span mask specially designed for FQN inference performs significantly better.

\vspace{1mm}
\noindent\fbox{
\begin{minipage}{8.2cm} \emph{Prompt learning with full span masking produces an effective code MLM for solving type inference as a fill-in-blank language task, which significantly outperforms the vanilla pre-trained code MLM.
The amount of prompt learning data does not significantly affect the model performance, but the masking strategy does if it does not fit with the data characteristics and task nature.} \end{minipage}}

\subsubsection{RQ2: Extrinsic Factors}
\label{sec:rq2}
This RQ includes the experiments on prompt similarity and API cardinality.
In this RQ, we use the model $M_F-3$ on the inference dataset.
We choose $M_F-3$ because the model performance becomes stable at $M_F$-3. 
Furthermore, $M_F$-3 is tuned with 30\% of source code files which give us more FQN prompts to obtain reliable results for prompt similarity analysis.

\paragraph{Prompt Similarity} \label{promp-similarity}
Our FQN prompts and code prompts are token sequences.
We remove mask tokens from the token sequences and convert them into bags of tokens.
We measure the similarity between a FQN prompt $P_{FQN}$ and a code prompt $P_c$ as $|P_{FQN} \cap P_c|$/$|P_c|$.
This token similarity measures how many FQN prompt tokens appears in the code prompt.
We compute token overlapping because names in code are meaningful for type inference.
Given a code prompt $P_c$ from the inference dataset, we compute its highest similarity with the FQN prompts in the prompt learning dataset.
We bin $P_c$ by similarity ranges and compute the type inference accuracy and BLEU-2 for these ranges.

\textbf{Result and analysis.}
Table~\ref{tab:similar-score} shows the results.
The similarities between FQN prompts and code prompts fall mainly in between 0.15 and 0.45, accounting for 72.5\% of code prompts.
4.6\% code prompts have (0, 0.15] similarities and 23.1\% code prompts have above 0.45 similarities with FQN prompts.
The model performance becomes stable as long as the code prompts have 0.35 similarity with FQN prompts.
Higher similarity increases the inference accuracy slightly (from 0.92 accuracy at (0.35, 0.45] to 0.92 accuracy at (0.65, 0.88]).
Even when code prompts have low similarity with FQN prompts (0.15 or below), the model still performs reasonably well (accuracy 0.76 and BLEU 0.90).
When the prompt similarity is $>$0.15, BLEU-2 scores fluctuate slightly around 0.94.

\begin{table}[]
    \centering
    \caption{Performance in Different Prompt Similarity Ranges}
    \begin{tabular}{|c|c|c|c|}
        \hline Similarity Range (\%) & Code Prompt (\%) & Accuracy & BLEU-2 \\\hline
        (0, 15] & 4.6 & 0.755 & 0.899\\\hline
        (15, 25] & 21.8 & 0.842 & 0.940\\\hline
        (25, 35] & 25.1 & 0.891 & 0.957\\\hline
        (35, 45] & 25.6 & 0.919 & 0.961\\\hline
        (45, 55] & 14.6 & 0.920 & 0.945 \\\hline
        (55, 65] & 6.3 & 0.918 & 0.937\\\hline
        (65, 88] & 2.2 & 0.924 & 0.956\\\hline
    \end{tabular}
    
    \label{tab:similar-score}
\end{table}

\paragraph{API Cardinality} \label{multiple-mapping}
When multiple APIs have the same simple name but different FQNs, the model must distinguish such APIs during type inference.
We bin the APIs by the number of APIs having the same simple name in different ranges, and compute the type inference accuracy and BLEU-2 for APIs in these ranges.

\begin{table}[]
    \centering
    \caption{Performance for Different API Cardinalities}
    \begin{tabular}{|c|c|c|c|}
        \hline Cardinality & Code Prompt (\%) & Accuracy & BLEU-2\\\hline
        1 & 34.7 & 0.925& 0.972\\\hline
        2 & 10.6 & 0.910& 0.961 \\\hline
        3 & 4.6 & 0.903& 0.957\\\hline
        4 & 3.9 & 0.900& 0.946\\\hline
        5 & 10.6 & 0.897& 0.955\\\hline
        (5, 10]  & 12.1 & 0.893& 0.950\\\hline
        (10, 20] & 5.4 & 0.915& 0.936\\\hline
        (20, 50] & 6.5 & 0.901& 0.910\\\hline
        (50, 100] & 2.9 & 0.892& 0.959\\\hline
        (100, 500] & 6.4 & 0.858& 0.931 \\\hline
        (500, 1000] & 1.7 & 0.842& 0.902\\\hline
        (1000, $+\infty$) & 1.7 & 0.826&0.874 \\\hline
    \end{tabular}
    
    \label{tab:Multiple-Mapping}
\end{table}

\textbf{Result and analysis.}
Table~\ref{tab:Multiple-Mapping} shows the results.
Only 34.7\% of the APIs in the inference dataset have unique simple name (i.e., cardinality=1).
Our model does not achieve 100\% accuracy for these unique-simple-name APIs because it sometimes generates inaccurate tokens.
For example, for the simple name \textit{BitString}, the correct FQN is \textit{org.apache.\underline{harmony}.security.asn1.BitString}.
Our model predicts an almost correct FQN \textit{org.apache.\underline{harm.}.security-.asn1.BitString} with just one wrong token (e.g., ``harm.'').

For the 65.3\% APIs with cardinality $>$ 1,
our model has very stable performance (accuracy 0.89-0.92 and BLEU-2 0.95-0.97) for the API cardinality up to 100.
The small fluctuation indicates that API cardinalities have only minor impact on model prediction.
For the extremely high API cardinality (500 and above), the mode performance degrades but is still acceptable (accuracy 0.82-0.84 and BLEU above 0.87-0.90).
Our inspection shows that many APIs with such extremely high API cardinalities are the Java APIs (e.g., \textit{toString()}, \textit{equals()}) or get/set methods (e.g., \textit{get()}).
Such APIs can appear in any classes, which makes it challenging to determine which specific class they come from based on only local context.

\vspace{1mm}
\noindent\fbox{
\begin{minipage}{8.2cm} \emph{The prompt-tuned code MLM performs accurately and stably for a wide range of prompt similarities and API cardinalities, and performs reasonably well even for very low prompt similarities (<0.15) or very high API cardinalities (>100).
This indicates that the model does not simply memorize what it sees during prompt learning, but acquire good inference capability for dealing with unseen FQNs, variant code contexts, and API ambiguity.} \end{minipage}}

\subsection{Practicality Evaluation}

\subsubsection{Motivation}
We investigates the practicality of our approach from three aspects.
First, we want to confirm how well our neural type inference approach performs on real partial code using certain library APIs, compared with existing dictionary-lookup based methods~\cite{Saifullah2019LearningFE,SnRConstraint,Phan2018StatisticalLO}
Second, in NLP studies~\cite{Brown2020LanguageMA,liu2021pre}, prompt-tuned models have shown strong inference capability for new tasks unseen during prompt learning. 
We want to investigate if prompt-tuned code MLM has similar capability for APIs and code contexts unseen during FQN prompt learning.
Third, we want to summarize failure patterns and plausible causes of our current approach to shed the light on improving neural type inference methods.

\subsubsection{Dataset and Metrics} 
In this study, we use two datasets of partial code collected from Stack Overflow.
The first dataset StatType-SO~\cite{Phan2018StatisticalLO} has been used in the experiments of previous type inference work~\cite{Phan2018StatisticalLO,Saifullah2019LearningFE,SnRConstraint}.
This dataset contains 268 partial code snippets.
Each code snippet is primarily about the use of APIs of one of the six libraries, Android, GWT, Hibernate, Joda Time, Xstream or JDK.
But many code snippets concerning the APIs of the first five libraries also use JDK APIs.
These code snippets contain 6-223 lines of code, with an average 28.
They involve 1838 and 2248 type inference points for non-FQN types and undeclared receiving objects, respectively, and include 1454 unique types to be inferred.
Furthermore, we build a new partial code dataset Short-SO from the Stack Overflow posts about the six libraries.
Furthermore, we build a new partial code dataset Short-SO from the Stack Overflow posts based on two criteria: 1) each code snippet must be related to some of the six libraries, and 2) contain no more than three lines of code.
We collect 20 code snippets for each library, i.e., 120 in total.
We use Short-SO to evaluate the capability of our model in face of little context information.
The code snippets in Short-SO involves 269 and 256 type inference points for non-FQN types and undeclared receiving objects, respectively, and include 286 unique types to be inferred.

We need to resolve the FQNs for non-FQN types and undeclared receiving objects in Short-SO for evaluation.
Based on the information in the posts (e.g., mentioned APIs, API links), two annotators (both are computer science graduate students) tried to make the code snippets in Short-SO compilable by manually fixing unresolved symbols in the code snippets.
One annotator performed the fixing and the other annotator validated the fixed code.
They discussed to resolve the disagreements.
We use accuracy (i.e., the strict ground-truth match) to measure the performance of our approach.
In all the experiments, we generate the code prompts with the all-unknown setting (see Section~\ref{sec:type inference}), in order to understand the lower bound performance of our approach in the worst-case scenario in the real world.
In practice, developers would likely know the FQNs of some simple names in a code snippet, which very likely helps to improve the prediction performance.
And we use $M_F$-3 as our model because the model performance becomes stable at $M_F$-3 (see Section~\ref{sec:rq1}).

\subsubsection{RQ3: Performance on Partial Code from Stack Overflow}
We run $M_F$-3 on both StatType-SO and Short-SO dataset.
We compare our approach with the state-of-the-art type inference tools for Java (i.e., COSTER~\cite{Saifullah2019LearningFE} and SnR~\cite{SnRConstraint}) as our baselines.

\textbf{Result and analysis.} 
Table~\ref{tab:StatType-SO} shows the results.
For StatType-SO, we compare the accuracy of our approach with the accuracy reported in~\cite{SnRConstraint}.
On average, our model $M_F$-3 performs better than both SnR and COSTER on StatType-SO (0.91 versus 0.87 and 0.71).
$M_F$-3 achieves much higher accuracy on GWT, JDK and Joda Time than both COSTER and SnR.
On Android, $M_F$-3 is close to SnR (0.91 versus 0.94), and is much better than COSTER (0.91 versus 0.43).
Our $M_F$-3 performs worse than COSTER and SnR on Hibernate and achieves only accuracy 0.76.
However, many errors are caused by the ground truth FQNs using the old package name (e.g., \textit{javax.persistence}) while the predicted FQNs using the new package name (e.g., \textit{jakarta.persistence}).
For the Xstream, our $M_F$-3 achieves the same performance as COSTER but is worse than SnR (0.88 vs. 1.00). 
Most errors in the Xstream are caused by the FQNs in the package \textit{com.cloudbees.api.config}.
Accidentally, none of the types in this package have been used in prompt learning, which means the model has zero knowledge of this package and its types. 
We further analyze this failure in Section~\ref{sec:rq5}.

$M_F$-3 achieves 0.99 accuracy on JDK partial code in StatType-SO and 0.97 accuracy on JDK partial code in Short-SO.
This is much higher than its performance on the JDK source code (0.84 accuracy).
This is because our effectiveness evaluation on JDK source code involve all kinds of JDK APIs with diverse forms of FQNs.
In contrast, the JDK APIs used in the partial code from Stack Overflow are mostly APIs from common packages (e.g., \textit{java.util}, \textit{java.io}) which are much easier to predict.

\begin{table}
    \centering
    \caption{Accuracy on StatType-SO and Short-SO}
    \vspace{-3mm}
    \begin{tabular}{|c|c|c|c|c|}
    \hline \multirow{2}*{Libraries}
    & \multicolumn{3}{|c|}{StatType-SO} & Short-SO \\
    
    \cline{2-5}
         ~ & COSTER & SnR & $M_{F}\!-\!3$ & $M_{F}\!-\!3$ \\\hline
        Android & 0.433 & 0.936 & 0.906& 0.819
 \\\hline
        GWT & 0.908 & 0.758 & 0.952 & 0.882 \\\hline
        Hibernate & 0.904 & 0.948 & 0.759 & 0.765 \\\hline
        JDK & 0.562 & 0.711 & 0.989 & 0.973\\\hline
        Joda Time & 0.571 & 0.895 & 0.975 & 0.971 \\\hline
        Xstream & 0.884 & 100.0 & 0.880 & 0.913 \\\hline
        Average accuracy & 0.710 & 0.875 & 0.910 & 0.887\\\hline
    \end{tabular}
    \label{tab:StatType-SO}
    \vspace{-3mm}
\end{table}

$M_F$-3 achieves high accuracy 0.89 on Short-SO.
Even though the code snippets in Short-SO is much shorter than those in StatType-SO, the performance of our $M_F$-3 does not differ much on the two datasets (0.91 versus 0.89).
For the four libraries Hibernate, JDK, Joda Time and Xstream, $M_F$-3 achieves very close accuracy on StatType-SO and Short-SO.
$M_F$-3 has relative worse accuracy for Android and GWT in Short-SO than StatType-SO. 
The main errors come from the wrong predictions for high cardinality methods, for example, \textit{getString()}, \textit{add()}.
In fact, our model relies on only limited local context (up to 5 code lines) for making inference.
Thus, the length of a code snippet does not matter.
Furthermore, our model makes inference over a neural knowledge base which is more powerful and flexible than the heuristic name and context matching over a symbolic knowledge base used in existing methods.
Saifullah et al.~\cite{Saifullah2019LearningFE} declared that ``the COSTER tool finds very few to no context in code snippets with only 1 or 2 lines of code and thus would fail to resolve FQNs of API elements''.

\begin{figure}
    \centering 
     \includegraphics[width=0.47\textwidth]{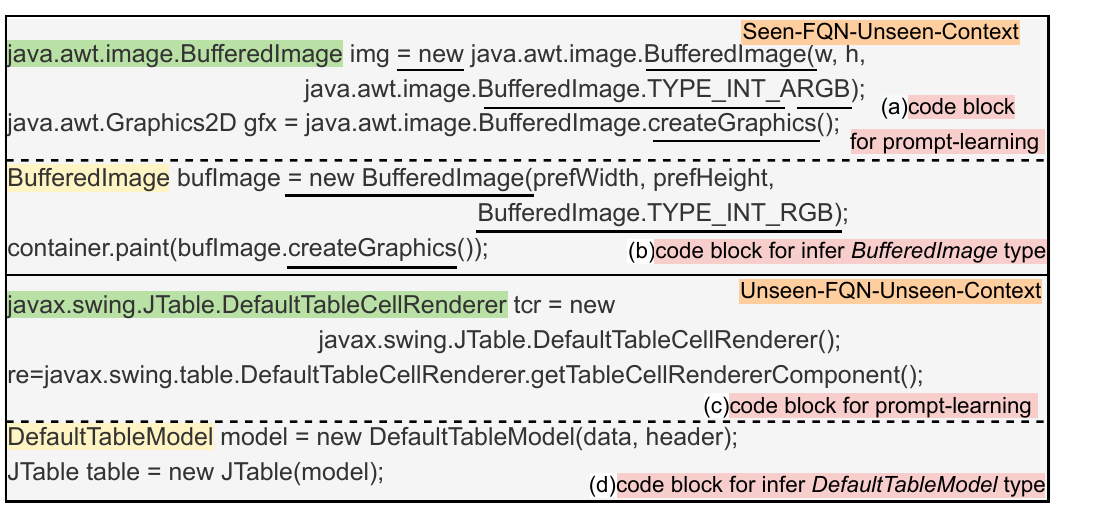}
    \caption{Example for Seen-FQN-Unseen-context and Unseen-FQN-Unseen-Context}
    \label{fig:Un/seen-FQN}
\end{figure}

\subsubsection{RQ4: Prediction Capability for Unseen APIs and Code Contexts}\label{exper:unseen-see}
An API in the SO datasets is seen if its FQN appears in the prompt learning data. 
Otherwise, the API in the SO datasets is unseen.
We have 3024 seen APIs and 1371 unseen APIs in the two SO datasets whose FQNs need to be predicted.
Otherwise, the code prompt is unseen.
We compute the prompt similarity as describe in Section~\ref{sec:rq2}.
We use the threshold 0.35 based on the impact analysis of prompt similarity in Table~\ref{tab:similar-score}.
We have 2626 seen code prompts and 1985 unseen code prompts.
We obtain four combinations of seen/unseen API and seen/unseen context, and compute the inference accuracy by $M_F$-3 for each combination on the two SO datasets.

\begin{table}
    \centering
    \caption{Accuracy for Seen/Unseen APIs and Code Contexts}
    \begin{tabular}{|c|c|c|c|}
        \hline \diagbox{Code-Context}{API} & Seen & Unseen & Overall \\\hline
Seen & 0.983 & 0.753 & 0.922 \\\hline
Unseen & 0.976 & 0.671 & 0.874 \\\hline
Overall & 0.980 & 0.712	& 0.902 \\\hline
    \end{tabular}
    
    \label{tab:see-unsee}
\end{table}

\textbf{Result and analysis.} 
Table~\ref{tab:see-unsee} shows the results of the four combinations. 
$M_F$-3 achieves high accuracy 0.98 for Seen-FQN no matter code context is seen or unseen.
When the FQNs to be predicted are seen during prompt learning, the accuracy in unseen context is only marginally lower than that in seen context.
When an FQN is seen during prompt learning, inferring it in a code snippet can be thought of as a dictionary lookup in our neural knowledge base.
However, compared with the dictionary lookup in a symbolic knowledge base~\cite{Saifullah2019LearningFE,SnRConstraint,Phan2018StatisticalLO}, our neural model can better handle context variations.
For example, the code block (b) in Figure.~\ref{fig:Un/seen-FQN} for inferring ``BufferedImage'' is an unseen code context.
Even the most similar code block (a) used in prompt learning is not that similar to the code block (b).
However, the model manages to pick up the most important code tokens (e.g., those with underlines) and correctly infer the FQN of \textit{BufferedImage}.
Note that our model does not use any explicit AST and data flow information, but is purely based on neural embedding of code tokens.

When the FQNs to be predicted are unseen during prompt learning, 
$M_F$-3 still achieves 0.75 accuracy and 0.67 accuracy for Unseen-FQN-Seen-Context and Unseen-FQN-Unseen-Context, respectively.
This zero-shot FQN prediction capability is impossible for existing methods, as unseen FQNs or code contexts will simply result in the OOV failures when looking up the symbolic knowledge base.
Our model's capability for unseen APIs and unseen contexts comes from the code naturalness that the model captures from large pre-training code corpus and subsequent prompt learning.
By code naturalness, developers name code elements with meaningful names and package related types in the same name space.
Figure~\ref{fig:Un/seen-FQN}(c) and (d) shows an example of Unseen-FQN-Unseen-Context.
The API \textit{javax.swing.JTable.DefaultTableModel} to be inferred is unseen in prompt learning, but the other API \textit{javax.swing.JTable.DefaultTable-CellRender} is seen in prompt learning.
Although the code block (d) for inference is not very similar to the code block (c) for prompt learning, the model could still see that both of them are somehow about \textit{JTable} and \textit{DefaultTable}, and make an intelligent guess of the FQN of \textit{DefaultTableModel}. 
This guess relies on not only the tokens visible in the prompts, but also the broader knowledge related to \textit{JTable} and \textit{DefaultTable} learned from the large code corpus.

\subsubsection{RQ5: Failure analysis}\label{sec:rq5}
We analyze common failure cases in the $M_F$-3's prediction results.
We count the number of errors $M_F$-3 makes for the FQNs in each package.
We analyze the top ranked packages with the most errors for each library.
For each such package, we collect distinct ground-truth FQNs with wrong predictions and count the times of wrong predictions for them.

\begin{table}
    \centering
    \caption{Frequent Wrong FQN Predictions}
    \begin{tabular}{|c|c|c|c|}
        \hline
        \makecell[c]{Library} & \makecell[c]{Top-1 \\ Package \\ Name (PN)} & \makecell[c]{\#DistGTFQNs \\ ($C_{PK}$/$C_{Library}$) } & 
        \makecell[c]{The Ground \\Truth FQNs \\ to be Inferred} \\\hline
    Joda Time & java.sql & 1 (6/13) & PN.Date \\\hline
 JDK & javax.swing & 1 (3/9) & \makecell[c]{PN.Swing-\\Worker} \\\hline
 Android & android.os & 1 (26/91) & PN.Bundle \\\hline 
 Xstream & \makecell[c]{com.cloudbees.\\api.config} & 2 (17/81) & \makecell[c]{PN.Enviro- \\nment, etc.} \\\hline 
 GWT & \makecell[c]{com.extjs.gxt. \\ui. client.widget} & 3 (11/50) & \makecell[c]{PN.Vertical- \\Panel, etc.}\\\hline
 Hibernate & javax.persistence & 17 (111/216)  & PN.Id, etc. \\\hline 
    \end{tabular}
    \begin{tablenotes}
    \centering
    \footnotesize
    \item\noindent $ C_{PK}$ and $C_{Library}$: the number of wrong predictions in a package and a library.
    \end{tablenotes}
    \label{tab:error-dis}
    \vspace{-5mm}
\end{table}

\textbf{Result and analysis.}
Table~\ref{tab:error-dis} shows the results.
Due to space limitation, we show only the top-1 package with the most wrong predictions (top-3 results in replication package).
Take Joda Time as an example.
The top-1 package is \textit{java.sql} which has one ground-truth FQN (\#DistGTFQN) (i.e., \textit{java.sql.Date}) been wrongly predicted as \textit{java.util.Date} for 6 times ($C_{PK}$).
These 6 wrong predictions accounts for 46.2\% of all errors ($C_{Library}$=13) for Joda Time.
Overall, the number of the distinct ground-truth FQNs with wrong predictions in a package is (much) smaller than the total number of errors in the package.
This mean there are many repeated errors for the same FQN to be inferred.
Our analysis identifies three major causes.


First, the errors are caused by similar usage contexts between the to-be-inferred FQN and the inferred FQN and the two FQNs are similar at the token level.
For example, $M_F$-3 repeatedly predicts \textit{java.util.Date} for \textit{Date} in the Joda Time code, but the ground truth is \textit{java.sql.Date}.
In fact, \textit{java.sql.Date} is the subclass of \textit{java.util.Date}, and the two classes have similar usage context.
We find that if the code context for \textit{Date} contains the SQL-related information (e.g., \textit{SQLException}), $M_F$-3 can correctly predict \textit{java.sql.Date}.
Otherwise, it prefers \textit{java.util.Date} which is applicable in generic contexts.

Second, the package name of the to-be-inferred FQNs does not appear in the prompt learning data.
That is, none of the types in this package is used for prompt learning.
We refer to this situation as zero-shot at package level, which is a very challenging task for the model.
For example, $M_F$-3 makes 17 wrong inferences for the two FQNs in the package \textit{com.cloudbees.api.config} of Xsteam, accouting for 21\% of all errors for Xsteam.
All these 17 errors are zero-shot failures.
The same situation for the 11 wrong inferences for the three FQNs in the package \textit{com.extjs.gxt.ui.client.widge} of GWT, accounting for 22\% of all errors for GWT. 
These errors could be reduced by using some types in the package to prompt the model.

Third, the inferred FQNs are actually correct but are from a different library version from that of the to-be-inferred FQNs.
That is, the inferred FQNs are false negatives.
In Hibernate, 111 errors (54\% of all errors for Hibernate and 24\% for all six libraries) are for the 17 to-be-inferred FQNs in \textit{javax.persistence}.
We obtain the ground truth FQNs in the Stack Overflow code snippets based primarily on the information in the post.
However, some of the FQNs in the code snippet can be replaced by the newer versions of the FQNs.
For example, the code snippet\footnote{https://stackoverflow.com/questions/3325387} is selected for Hibernate.
We obtain the package name \textit{javax.persistence} of the API types (e.g., \textit{ManyToOne}, \textit{Column}).
However, $M_F$-3 predicts \textit{jakarta.persistence} for these API types.
In fact, they are the same APIs in different version of, i.e., \textit{javax.persistence} was renamed to  \textit{jakarta.persistence} in JPA 3.0.
As we use the JPA 3.1.0 to prompt-tune the model, the model makes inference by the newer version of APIs.

\noindent\fbox{
\begin{minipage}{8.2cm} \emph{Our model achieves higher type inference accuracy than the state-of-the-art dictionary-lookup methods~\cite{Saifullah2019LearningFE, Phan2018StatisticalLO, SnRConstraint}, and it achieves unprecedented results for short code snippets and unseen FQNs and code contexts. Our model may make mistakes for APIs with similar functionalities and in zero-shot package setting. 24\% wrong inferences are false negatives as the inferred FQNs and the ground-truth FQNs are the same APIs from different library versions.} \end{minipage}}

\section{DISCUSSION}

This section summarizes our experience, envision future directions, and discuss threats to validity.
\vspace{-3mm}
\subsection{Model Enhancements and Extensions}
Neural type inference is a completely new approach.
From our experience in this study, we highlight research points that are worth further investigation for enhancing neural type inference methods.

\subsubsection{Pre-training Code MLM}
We currently use the pre-trained CodeBert~\cite{Feng2020CodeBERTAP}.
This allows us to study the prompt learning method for type inference on top of a well studied, mature code MLM.
CodeBert has only 1.25 millions parameters, which is much smaller than the latest pre-trained natural language model and code model, for example, GPT3\cite{Brown2020LanguageMA} and its code variant CoPilot~\cite{Pearce2021AnEC} which have 175 billions parameters.
We are investigating if larger pre-trained model may lead to better performance.
As code MLM is pre-trained on raw source code, the only technical limitation to train larger code MLM is computing resources.
An important hyperparameter of MLM is the context window size.
Our model uses 512 tokens.
This fits well with short partial code and reduces irrelevant contextual noise, but it limits the number of contextual code lines in each code prompt.
Through the design of FQN prompts with overlapping code lines, our model can propagate long-range contextual information.
The state-of-the-art MLM uses much larger context window (e.g., 2048)~\cite{gao2020pile}.
It is worth investigating if larger context window (i.e., more contextual code lines in a code prompt) would allow the model to better capture long-range contextual information or it may introduce more noise which may interfering the inference.


\subsubsection{Prompt Learning}

Due to random sampling, we accidentally encounter the cases in which none of the types in a package are sampled for prompt learning (e.g., \textit{com.cloudbees.api.config} in Xstream).
As the model has zero knowledge of this package, it fails to infer the FQNs of the types in this package.
This result suggests our model have good few-short learning capability, but zero-shot at package level is still very challenging to it.
As such, a better prompt learning strategy should expose the model to as many package prefixes as possible and let the model see some FQNs from each package which would maximize the model performance.
The FQN prompts in this work mark only the FQNs in the focused code line but keep the contextual code lines unchanged.
We were wondering if we could make the learning more challenging by masking some FQNs and/or other tokens in the contextual code lines.
We hypothesize this could force the model to better learn the correlations between the focused code and the contextual code.
This leads to another interesting question: how many contextual tokens should we mark?
The original BERT marks 15\% of tokens.
Later studies~\cite{He2021MaskedAA} show higher mark ratio (e.g., 75\%) may results in better results.
We need to identify a good trade-off point at which there are sufficient masks to challenge the model to learn harder, but masks are not too many so that no sufficient information for the model to learn.


\subsubsection{Type Inference}
We make type inference at each inference point independently.
The benefit is that it removes the reliance on code analysis.
However, type inference points may be correlated (e.g., the same type of different variables, the same variable used at different places).
When the correlated places are inferred simultaneously, the generated code prompt will contain some FQNs, which helps the model subsequently infer other FQNs.
Such code prompt can be considered to lie in between the leave-out-one setting and the all-unknown setting in our experiments.
It is worth investigating how this type of code prompts affects the performance of the model on type inference.
However, this would require the design of corresponding FQN prompt and masking strategy.
A concern is simultaneous multiple type inference would increase the reliance on code analysis.
In addition, in the effectiveness and practicality evaluation, we generated the code prompts of the upper bound on Github datasets, and the ones of the lower bound on Short-SO.
In the future, we will generate the code prompts of the lower bound on the GitHub datasets, to investigate the performance of the model.
Our model generates a small number of incorrect FQN formats (e.g., two consecutive ``.'').
Such format errors can be avoided by post-processing the language model type inference results with code syntax~\cite{wang2021code}. More interestingly, this inspire us whether reinforcement learning (RL) could be useful in neural type inference methods.
We could encode some FQN format and syntactic rules as the RL rewards and let RL guide the MLM to avoid incorrect FQN formats or other FQN-related syntactic errors.



\subsubsection{Type Inference for Other Programming Languages (PLs)}\label{sec:otherpl}
In this work, we focus on statically typed PL and evaluate our approach on Java.
As our approach has little reliance on code compilation, there are no major barriers to extend our model to other statically typed code, such as C\#.
In fact, CodeBert is trained with a code corpus mixing six programming languages.
Many PLs use the same FQN form (\textit{a.b.c}), while some PLs use other forms (e.g., \textit{a::b::c}).
Furthermore, different PLs may have different naming conventions.
An interesting question is whether a PL-hybrid type inference model can provide an one-for-all solution, or we need PL-specific model because different PLs may interfere with each other.
More challengingly, how can neural type inference methods be applied to dynamically-typed PLs?
From our experience, we first need to expose the MLM to as much code information as possible during pre-training.
Then, we need to prepare some FQN-annotated code examples that have a good coverage of typical code use for prompt learning.
As prompt learning has few shot learning capability, this preparation step (even manually) could still be feasible.

\vspace{-3mm}
\subsection{Downstream Applications}
Making effective use of online code examples has been a long-term interest in software engineering~\cite{zhang2019analyzing,umarji2008archetypal,inproceedings, brandt2009two,baltes2019usage,wu2019developers}.
Resolving types in partial code is the first step towards this goal.
Different from existing type inference methods~\cite{Phan2018StatisticalLO,Saifullah2019LearningFE,SnRConstraint}, our approach uses a much more lightweight method to construct an easily-extensible neural knowledge base of APIs and usage contexts.
This opens the door to many useful downstream applications.
For example, we could use our model to parse code examples in API documentation to enrich the code examples in software knowledge graph~\cite{Li2018ImprovingAC, Sun2019KnowHowIP, Liu2019GeneratingQC} with API type information. 
As another example, existing code search is still mainly based on keyword matching, because advanced code search methods~\cite{linstead2009sourcerer,8453174,9054830,huang2019enhance,huang2019qe} all more or less rely on code analysis which makes them hard to deploy.
Our model could lower the barrier to deploy these advanced code search engine by annotating large code base with API types to support API-centric code search.
Last but not least, our model could be integrated in Stack Overflow or IDE as a neural compiler, which can provide just-in-time type fixing assistance when developers read code snippets on Stack Overflow or copy code snippets to the IDE from the Web.
For example, IntelliJ IDE marks ``util'' in ``util.List'' as an unresolved symbol but cannot offer any fixing suggestion.
Our model can easily fix this unresolved symbol as ``java.util.List''.

\vspace{-3mm}
\subsection{Threats to Validity}

As a deep learning based method, data errors is a major threat.
We validate our data processing modules, and examine many inputs, intermediate results and outputs to ensure the data correctness.
The FQNs in the StatType-SO dataset come from previous studies~\cite{Phan2018StatisticalLO}, and we double-checked their validity.
For our own Short-SO dataset, two authors labeled the ground-truth FQNs.
We also test our tool chain to ensure its correctness.
Note that our approach treats source code as text and use mature NLP tools (e.g., WordPiece~\cite{Wu2016GooglesNM} for tokenization).
Our approach does not require any sophisticated program analysis, except for obtaining FQNs from library source code using mature Java parser~\cite{Pawlak2016SPOONAL}.
The major external threat is that our approach was evaluated on only six Java libraries.
We use these six libraries because they were used in all recent type inference studies~\cite{Phan2018StatisticalLO,Saifullah2019LearningFE,SnRConstraint}.
The six libraries cover two major SDKs (JDK and Android) and other specific application domains including XML parsing (Xstream), web tookit (GWT), date and time (Joda Time), database (Hibernate).
These SDKs and libraries have different naming conventions and FQN characteristics.
We will apply our approach to more Java libraries to further confirm its generalizability.
We will also extend our approach to other statically-typed program languages (e.g., C\#, Rust, Swift).
As our approach treats code as text, the extension should be straightforward.


\vspace{-3mm}
\section{Related Work}

Recently, the great success of large pre-trained language models (PLM) (e.g., BERT~\cite{Lewis2020BARTDS,Heinzerling2021LanguageMA}, RoBERTa~\cite{liu2019roberta}, GPT-3~\cite{Brown2020LanguageMA}, T5~\cite{Raffel2020ExploringTL}) in a variety of NLP tasks, spawns a surge of code PLMs (e.g., CuBert~\cite{kanade2020learning}, CodeBERT~\cite{Feng2020CodeBERTAP}, CodeT5~\cite{wang2021codet5}, CoPilot~\cite{Pearce2021AnEC}). 
The application of language models on code is built on source code naturalness~\cite{Devanbu2012OnTN,Allamanis2018ASO}.
As the pre-training tasks (e.g., mask prediction) are agnostic of downstream tasks, studies~\cite{Paradigm} show the vanilla output (e.g, MLM head) of code PLMs does not encode much code syntax and semantics.
Through external probing, recent works~\cite{Wan2022WhatDT, Troshin2022ProbingPM, Buratti2020ExploringSN} show code PLM contain information about code syntax, the notions of identifiers and namespaces, and natural language naming. 

To make use of the rich code information in code PLMs in downstream software engineering tasks, all existing work follows a ``pre-train, finetune'' paradigm which takes the PLM as a feature extractor and connect it to a downstream neural network (e.g., a Multi-Layer Perceptron~\cite{Vaswani2017AttentionIA}).
Then, the PLM, together with this neural network, can be trained on a small and specialized data for downstream tasks (e.g., code search~\cite{Husain2019CodeSearchNetCE}, AST tagging~\cite{wan2022they}, vulnerability prediction~
\cite{morrison2015challenges}, clone detection~\cite{wang2020detecting}, code summarization~\cite{ahmad2020transformer})~\cite{Karmakar2021WhatDP,troshin2022probing,Wan2022WhatDT}.
In this paradigm, the PLM parameters can be fixed or fine-tuned (usually the last few layers~\cite{zhou2019improving}), and the task-specific learning occurs mainly in the downstream heterogeneous neural network.
We refer to this paradigm as outsourcing upgrade.
In contrast, we adopt a completely new paradigm ``pre-train, prompt and predict'' which is a self-upgrade of pre-trained code MLM, in the sense that model pre-training and prompt learning are homogeneous at the same MLM core.
Due to this alignment, prompt learning efficiently stimulates task-agnostic pre-trained code MLM to recognize FQN syntax and usage with some FQN prompts from library code.

Studies~\cite{Raffel2020ExploringTL,Brown2020LanguageMA,Schick2021ItsNJ, Schick2021ExploitingCF} show the meta-learning capability of large PMLs: the model conditioned on a task description and some examples of the task can learn to complete new instances of the task.
Early methods focused on hand-crafted prompts~\cite{Raffel2020ExploringTL,radford2019language,Schick2021ItsNJ, Schick2021ExploitingCF}, and the effects are sensitive to prompt variants~\cite{shin2020autoprompt}.
Subsequent work proposes to automate the search of prompts, including both discrete prompts~\cite{Gao2021MakingPL,shin2020autoprompt} and continuous prompts~\cite{Lester2021ThePO,Li2021PrefixTuningOC,Tang2022ContextTuningLC}.
Natural language prompts cannot be directly applied in our task, because they are uniform for all inputs. 
In light of the variable-length form of FQNs and variant API usage context, we design automatic and contextual FQN prompts which differ conceptually from existing NLP prompt designs.
Furthermore, our FQN prompts use full span masks, instead of random masking commonly used in NLP tasks.

Several NLP studies~\cite{Roberts2020HowMK,petroni2019language, Jiang2020HowCW, heinzerling2020language} show that PLMs can serve as neural knowledge bases of real-word entities and relations.
Compared with symbolic knowledge bases, neural knowledge bases do not require schema engineering and human annotations, and support an open set of queries.
Our prompt-tuned code MLM is a neural knowledge base that packs API names and code contexts in large code corpus into model parameters.
This is completely different from the symbolic knowledge bases that explicitly maps API names and code contexts in existing type inference methods~\cite{Saifullah2019LearningFE, Phan2018StatisticalLO, SnRConstraint}, require to compile and parse a large number of software projects using libraries APIs.
Furthermore, we formulate type inference as a fill-in-blank language task, which aligns perfectly with the learning objective of prompt-tuned code MLM.
As such, our type inference retrieves the FQN knowledge from the model through neuron activation, as opposed to symbolic name and code-context matching.
This neural type inference supports fuzzy reasoning of the correlations between API names and code contexts, which may not be explicit in the partial code under analysis.

\section{Conclusion  and Future Work}

This paper presents the first neural type inference method built on a novel ``pre-train, prompt and predict'' paradigm.
Our approach solves type inference as a fill-in-blank language task using a prompt-tuned code MLM, thus removing the reliance on compiling a large number of software projects to construct symbolic API knowledge base.
Our approach perfectly aligns model pre-training, prompt learning and inference with the same mask learning objective.
We design novel automatic and contextual FQN prompt, full-span mask strategy for prompt learning, and variable-length mask prediction method for FQN inference.
Our experiments show prompt learning takes effect with as minimum as 10\% of library source code files, and our model is robust in handling unseen FQNs, code context variations and API ambiguities.
Our model achieves excellent performance for short partial code with little context information, and the achieves promising capability of few-shot type inference. In the future, we will explore ways to enhance and extend our approach and develop novel applications for partial code search and reuse enabled by our model's unprecedented type inference capability.




\section{Acknowledgements}
The work is partly supported by the National Nature Science Foundation of China under Grant (Nos. 61902162, 61862033), the Nature Science Foundation of Jiangxi Province (20202BAB202015), Postgraduate Innovation Fund Project of Jiangxi Province(YC2021-S308), and the Science and technology Key project of Education Department of Jiangxi Province (GJJ210307).

\balance
\bibliography{sample}

\end{document}